\documentstyle{l-aa}

\input psfig  
\begin{document}

\thesaurus{02.18.6, 03.13.7, 11.03.1, 12.12.1,  13.25.3}

\title{An XMM pre-view of the cosmic network: galaxy groups and filaments}

\author{
M. Pierre\inst{1}
\and G. Bryan\inst{2,3}
\and R. Gastaud\inst{1}
}
 
\offprints{ M. Pierre    (mpierre@cea.fr)}

\institute{
CEA/DSM/DAPNIA, Service d'Astrophysique, F-91191 Gif sur Yvette, France
\and  Physics Department, MIT, Cambridge, MA 02139, USA
\and  Hubble Fellow
 }

\date{submitted}
\maketitle

\begin{abstract}

A large fraction of the baryons today are predicted to be in hot,
filamentary gas, which has yet to be detected.  In this paper, we use
numerical simulations of dark matter and gas to determine if these
filaments and groups of galaxy will be observable be XMM.  The
simulated maps include free-free and line emission, galactic
absorption, the XMM response function, photon noise, the extragalactic
source distribution, and vignetting.  A number of cosmological models
are examined as well as a range of very simple prescriptions to
account for the effect of supernovae feedback (preheating).  
We show that XMM has a good chance of
observing emission from strong filaments at $z\sim 0.1$.  This becomes much
more difficult by $z \sim 0.5$.
The primary difficulties lie in detecting such a large, diffuse object and in selecting an appropriate field.
We also describe the range of group sizes that XMM should be sensitive
to (both for detection and spectral analysis), although this is more
dependent on the unknown nature of the feedback.  Such observations
will greatly improve our understanding of feedback and should also
provide stronger cosmological constraints.


\end{abstract}

\section {Introduction}

The filamentary large scale distribution of matter in the universe is
a consequence of the gravitational instability of a medium that was
relatively smooth in early times, with the non-linear patterns we see
today having evolved from small density fluctuations.  There have been
two competing stories for the formation of structures: the (Russian)
pancaking picture (\cite{zel70}) and the (Western) hierarchical
clustering picture (\cite{pee80}). The ``cosmic web'' paradigm (\cite{bon96}) 
reconciles these two scenarios, showing that the final-state is
actually present in embryonic form in the overdensity pattern of the
initial fluctuations, with nonlinear dynamics just sharpening the
image. This has several immediate applications to observations, in
particular: superclusters are predominantly cluster-cluster bridges
and the strongest filaments exist between aligned clusters.

The study of galaxy clusters has undergone spectacular developments in
the last decade, ranging from the galaxy content to the detailed
physics of the hot intra-cluster gas (ICM) and also addressing the
question of the large scale distribution of clusters. Conversely,
intermediate density structures, that we would define as density
contrasts of the order of $\delta \rho / \rho \sim 4$, i.e. loose
galaxy groups and filaments, have captured less attention so far. The
main reason being their comparatively humble appearance in the optical
or X-ray sky.

However, both theory and simulations indicate that a better knowledge
of the filaments will improve our understanding of cosmology.  Many
current models suggest that a large fraction of the baryons at low
redshift should be in filaments (e.g. \cite{mir96}; \cite{cen99}).
Heated to a temperature intermediate between that of hot clusters and
cool voids, this would imply that presently most of the baryons remain
undetected.  Moreover, the overall cosmic network - best underlined by
the numerous groups and filaments - may show different topologies
depending on the value of the cosmological parameters and the nature
of dark matter.

Filaments and groups are superb laboratories for studying the effects
of galaxies on their environment and, in turn, the impact of this
environment on galaxy evolution.  The velocity dispersion of groups
and filaments ($\sim$ 400 km/s) is larger than the internal
velocities of galaxies themselves, thus interactions are frequent and
the intragroup gas should have very similar properties to the
neighbouring filamentary gas. Feedback from galaxy/star formation is
expected to play a major role --- much more, probably, than for
massive clusters where gravitational physics may be sufficient to
explain the bulk of their properties.  Also, many quasars are thought
to be in groups and may have significant effects on their environment.
In the various observed X-ray/optical correlations, groups tend to
show a larger dispersion than clusters (e.g. \cite{mul98}).  There are
also some systematic difference in the general properties of clusters
and groups (\cite{ren97}).  These results certainly suggest that galaxies
play an important role in groups and filaments, but this process is
very poorly understood.
This is one of the compelling
arguments in favor of deep and extended X-ray surveys: a large sample
of groups (free of projection effects or of any optical selection
biases, which can be very stringent e.g. ``compact'' groups) would
enable the statistical study of their temperatures and abundances to
be further compared with their optical properties and theoretical
predictions. Since groups occupy the intermediate range in the cosmic
mass function --- between clusters and galaxies --- a better
understanding of their global properties would provide an alternative
view of the missing link between quasi-linear and extremely non-linear
systems.

Our observational knowledge of the truly filamentary medium is even
slimmer than for groups, bordering on non-existent. Attempts to detect
the ``supercluster'' or filament gas directly from its bremsstrahlung
emission have led to very marginal or negative results so far
(\cite{bri95}; \cite{wan97}; \cite{kul99}; see also
\cite{mar99}). Although the temperature of this gas is expected to be
around 1 keV, its low density considerably limits its
potential emissivity in the X-ray or EUV band.  It is thus primarily a
technical challenge to observe this warm component which is predicted
to account for some 20-40\% of the total baryonic mass by redshift $z=0$ (\cite{cen99}).
In addition to a direct window into the physics of this medium,
closely connected with the cooler components such as Ly$\alpha$ clouds
and galaxies, quantitative observations would put an upper limit to
the contribution of filaments to the auto-correlation function of the
X-ray background.
 
Thanks to its unrivaled sensitivity, XMM\footnote{http://astro.estec.esa.nl/XMM/xmm.html} is to open a new era in the
[0.1-10] keV range. One aspect of the mission is focussed on high
resolution spectroscopy, the other main one being the observation of
faint extended diffuse emission. In the EPIC imaging mode at 1 keV,
XMM has a Half Power Diameter of 13" on-axis, a collecting area of
2600 cm$^2$ and an energy resolution of 20\%. Beside the detailed
study of massive clusters of galaxies, XMM is thus ideally suited to
search for low surface brightness extended structures such as very
distant galaxy clusters (at redshifts substantially greater than one,
if they exist!) or groups and filaments out to a redshift close to
unity.

XMM instruments are now fully calibrated and the launch is scheduled
for the  end of the millennium.  It is thus timely to
investigate carefully which of the issues related to intermediate
density structures can be efficiently addressed by XMM. This will have
a direct impact on the preparation of dedicated observations as well
on their analysis. In order to provide quantitative arguments, we have
simulated XMM images.  The X-ray emissivity of a filamentary
region at a range of redshifts was computed from high resolution
hydrodynamical simulations combined with a detailed plasma code and
folded with the XMM spectral and imaging responses. We present here
the XMM simulated images and further discuss the physics that could be
extracted from them.

The paper is organized as follows. In the next section, we present the
hydrodynamical code and the various cosmological models and epochs
that were chosen for the the simulations. Section 3 describes how the
X-ray images were produced. The general discussion on the
detectability of groups and filaments is given in Section 4 as well as
practical hints for future large surveys of groups.  In the conclusion
we present further observational prospects in connection with other
wavelengths.

\section{The AMR simulations}

\subsection{Basic ingredients of the simulations}

In order to generate realistic XMM images, we have performed a series
of cosmological simulations incorporating both dark matter and
baryonic gas.  The method used is an adaptive mesh refinement (AMR)
algorithm that models the dark matter with particles and the gas with
a series of meshes.  More complete descriptions of the method are
available elsewhere (\cite{bry97}; \cite{nor99}), but we briefly
outline it here.  An initial grid was set up at high redshift with
density and velocity fields drawn from a power spectrum appropriate
for the cosmology of interest.  The particles were initialized using
the Zel'dovich approximation.  The equations of hydrodynamics were
solved on the mesh using the piecewise parabolic method (\cite{col84};
\cite{bry95}).  Since the mesh is fixed, gravitational collapse
inevitably causes structures to shrink to sizes comparable to the mesh
spacing, causing inaccuracies.  AMR solves this problem by identifying
regions which need enhanced resolution and laying down a finer mesh
over those volumes.  The solution is interpolated onto this finer grid
--- which typically has cell sizes half as large --- and an improved
solution is computed there.  The process can be repeated as necessary
and a hierarchy of grids is generated, with cell sizes varying as
small as necessary (up to a predefined limit in order to limit the cpu
time required).

In Table~\ref{tab:cosmo}, we list the parameters of the simulations
performed.  They are drawn from three different cosmologies: 1) a
standard cold dark matter (SCDM) model, in order to link with previous
work on this model, 2) a cosmological constant-dominated universe
($\Lambda$CDM), which agrees with the majority of current
observational data, and 3) a tilted-CDM (tCDM) model.  Together, these
three choices should provide a good range in the currently viable
parameter space.

\begin{figure}
\psfig{file=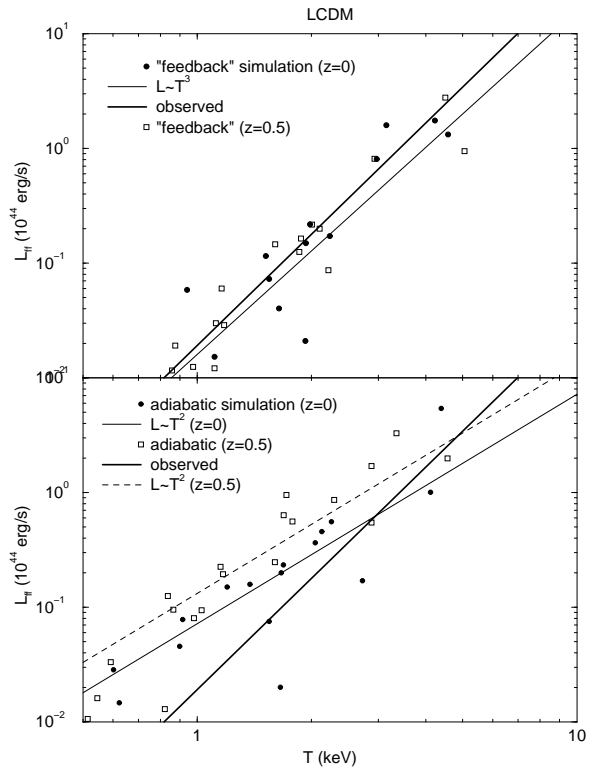,width=9cm}
\caption[] { 
The introduction of ``mild'' feedback (see text) in the simulations
results in a decrease of the luminosity of small groups and,
consequently, steepens the $L_X-T$ relationship in agreement with the
observations: observations give $\sim L_X\propto T^{3}$ whereas
$L_X\propto T^{2}$ is expected in the case of a pure gravitational
collapse. The observed correlation appears to be very tight when
``cooling flow'' clusters are excluded (\cite{AE99}) and shows no
obvious evolution out to $z \sim 0.4$ (\cite{MS97})}
\label{L-T}
\end{figure}

In addition to changing the cosmology, we also experimented with
additional physical processes in the simulations.  In some simulations
we included a simple equilibrium cooling-curve (\cite{sar87})
(although we should note that the spatial resolution of these
simulations, 50 $h^{-1}$ kpc, were not sufficient to fully resolve the
cooling instability).  The physical state of the intergalactic medium
is intimately linked with galaxy formation (through galactic outflows)
and hence star formation and supernovae.  In fact, simulations of
clusters (e.g. \cite{nav96}; \cite{bry98}) show that gravity and
adiabatic gas physics alone do not correctly predict the
luminosity-temperature relationship.  Instead of the observed relation
which is approximately $L_X \sim T^3$, they predict $L_X \sim T^2$.
The difference is most likely due to feedback from stars
(\cite{kai91}; \cite{bal99}), which those simulations did not include.
We would like to take into account the effect of feedback here,
unfortunately the input of energy from supernovae and stellar winds is
technically difficult to include properly in simulations
(\cite{met94}).  In place of a full treatment, we attempt to include
this effect by pre-heating the gas, effectively setting a minimum
entropy (see also Navarro, White and Frenk 1996).  This does not have
a large effect on the hot gas in the outskirts of clusters which, due
to shocking, naturally has a high entropy.  However, it does produce a
larger core radius, decreasing the central gas density compared to a
simulation without pre-heating.  This decreases the luminosity of
small clusters and groups (since the X-ray emissivity is proportional
to the square of the density), and can reproduce the observed
luminosity-temperature relationship.  Since it is not clear what this
initial entropy should be, we tried two cases: a mild feedback in
which the initial gas temperature is set to $4 \times 10^7$ K at
$z=20$ (and would adiabatically cool to $9 \times 10^4$ K by $z=0$ for
a fluid element which follows the cosmic mean density without
shocking), and a more extreme version in which the initial temperature
is 5 times higher.  The impact on the $L_X-T$ relationship of
introducing galaxy feedback in the simulations is shown in
Fig. \ref{L-T}.  This form of ``feedback'' is successful in
reproducing the observed relation.  Clearly, a simple pre-heating of
the gas does not produce the same energy distribution as the feedback
from millions of massive stars; however the extremes adopted here (no
feedback at all contrasted with heating the gas everywhere) will at
least indicate which diagnostics are sensitive to the feedback and by
how much.

\begin{figure}
\psfig{file=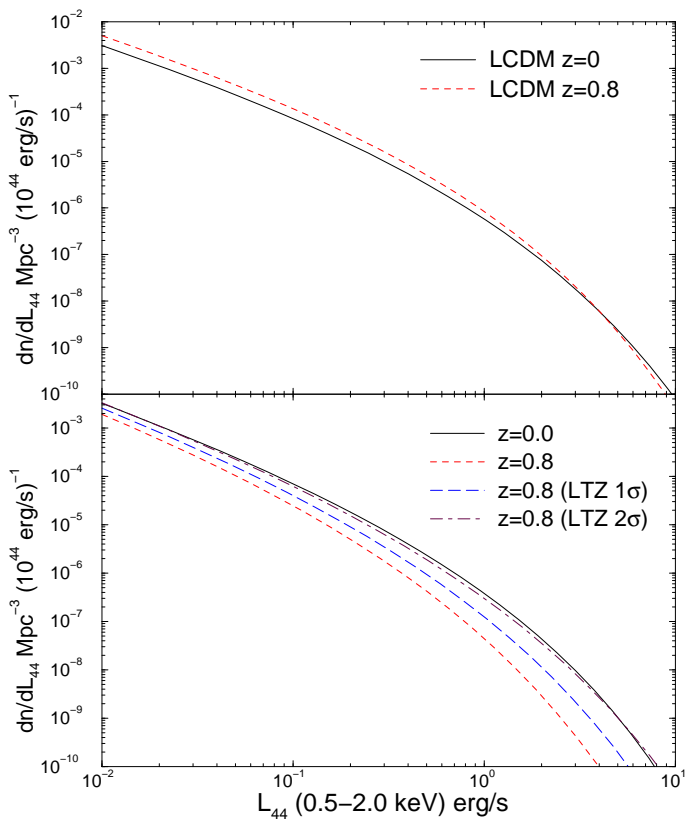,width=9cm}
\caption[] {The predicted luminosity function of the $\Lambda$CDM (top) and
tCDM (bottom) models at $z=0$ and $z=0.8$.  In the bottom panel, we
also show the effect of varying the amount of redshift evolution in
the observed $L_X-T$ relation by one and two sigma.  Since observations
seem to indicate little or no evolution in the luminosity function,
the tCDM model is marginally inconsistent.}
\label{fig:lum_function}
\end{figure}

The cosmological constant-dominated model we have selected is in good
agreement with most available observational data, and in particular
with the shape and evolution of the luminosity function.  The flat,
tCDM model is certainly less-favoured by the present data.  It is
consistent with the number-density of clusters and COBE measurements
of the microwave background; however, it does not match with the many
indications of a low-density universe, nor with the recent supernovae
data implying the presence of a cosmological constant.  We include it
to indicate the range of variation that cosmology will introduce in
our filament properties.  The X-ray properties of the models are
particularly important for this study and so in
Figure~\ref{fig:lum_function}, we plot the luminosity function for
each model, as predicted from a simple Press-Schechter prescription
combined with a relation between mass and temperature, coupled with
the observed luminosity-temperature relation (see \cite{bry98} for
more details on this procedure).  While both models agree at $z=0$,
observations are now becoming available at much higher redshifts.
For example, the lack of evolution out to $z \sim 0.8$ for moderate to
high-luminosity clusters (\cite{ros98}) places tighter constraints on
the models.  In Figure~\ref{fig:lum_function} we also plot the
predicted luminosity function at $z = 0.8$.  For the $\Lambda$CDM model, the
lack of evolution is naturally predicted; however for the tCDM model
to agree, we have to stretch the bounds of the observed evolution (or
rather lack of evolution) of the luminosity-temperature relation,
which is the most uncertain part of the modeling procedure.  For
example, using a very recent determination of the evolution of the
$L_X-T$ relation (\cite{rei99a}), the tCDM model disagrees at the 2
$\sigma$ level.  Note, however, that the observational luminosity
function at high redshift is still somewhat uncertain (see for example
\cite{rei99b}, where a deficit of very high luminosity clusters is
observed).

\begin{table*}
\caption[]{AMR simulation models - {\bf comoving cell size: (50~kpc/$h$)$^3$}}

\begin{tabular}{||l|c|c|c|c|c|c|c|c||}
\hline \hline
\label{tab:cosmo}
Model  &   $\sigma_{8}$ & $h$ & $\Omega_{matter}$ &
$\Omega_{\Lambda}$   & $\Omega_{b}$ & $n$ & cooling? & 
``feedback''?  \\ \hline 
CDM  & 0.95 & 0.6 & 0.4 & 0.6 & 0.06 & 1.0 & no  & no  \\ \hline
$\Lambda$CDMf1  & 0.95 & 0.6 & 0.4 & 0.6 & 0.06 & 1.0 & no  & mild  \\ \hline
$\Lambda$CDMc  & 0.95 & 0.6 & 0.4 & 0.6 & 0.06 & 1.0 & yes & no  \\ \hline
$\Lambda$CDMcf1  & 0.95 & 0.6 & 0.4 & 0.6 & 0.06 & 1.0 & yes & mild  \\ \hline
$\Lambda$CDMf2  & 0.95 & 0.6 & 0.4 & 0.6 & 0.06 & 1.0 & no & extreme  \\ \hline
tCDM          & 0.6  & 0.5 & 1.0 & 0.0 & 0.06 & 0.81 & no  & no  \\ \hline
SCDM          & 0.6  & 0.5 & 1.0 & 0.0 & 0.06 & 1.0 & no  & no  \\ 
\hline \hline
\end{tabular}  \\  
\end{table*}

\begin{table*} 
\caption[]{Steps in the simulation of the  XMM images }  

\begin{tabular} {||l|l ||}
\hline \hline
Step 	   &  Operation	   \\ \hline
{\bf 0a}         & AMR cube + plasma code $->$ X-ray emissivity cube   \\
{\bf 0b}         & {\bf 0a} + XMM response + Exp. time $->$ photon cube     \\ \hline 
{\bf 1a}         & projected photon cube   \\
{\bf 1b}         & {\bf 1b} + Poisson noise      \\ \hline
{\bf 2a}         & a single XMM field of view from {\bf 1a}   \\
{\bf 2b}         & {\bf 2a} + extragalactic point sources  \\ \hline \hline
\end{tabular}  \\ 
\label{tab:imasimu} 
\end{table*}    
 
\subsection{The extracted filament}

Since we are primarily interested in the filament environment, we
specifically selected such a region.  This was done by initially
simulating --- at low resolution --- a cube 100 $h^{-1}$Mpc on a side.
The output was examined, and a volume 18 $\times$ 23 $\times$ 33
$h^{-3}$ Mpc$^3$ was selected.  This region lay between two large
clusters and was the strongest filament in the box.  In order to
better sample the initial conditions, the simulation was rerun with an
addition mesh of size 28x34x50 placed over the region.  As many
particles as grid points were used.  The adaptive meshing was then
turned on and a relatively high resolution simulation of this region
was performed.  The rest of the 100 $h^{-1}$ Mpc cube was again
simulated at low resolution, which is sufficient to provide the
necessary gravitational tidal field.

\section{Simulations of the XMM images}

From the temperature and density cubes, we have calculated for each
simulation cell, the X-ray plasma emissivity using a redshifted
Raymond-Smith code having a metallicity of 0.3 solar. The choice of
this latter value - which is comparable to that of galaxy clusters,
although the total production of heavy elements by group galaxies is
certainly lower - has been guided by the fact that mixing and
enrichment are probably more efficient than in clusters, the medium
being much less dense. Then, assuming a given exposure time, we
computed the photon numbers to be received from each cell by XMM
taking into account the complete instrumental response (a composite
matrix made of the responses of the 2 MOS CCDs and of the PN CCD in
imaging mode folded with the X-ray telescope response). This was
done for the  [0.4-4] energy band, best suited for detecting structures having temperatures of the order of 1-2 keV; we do not consider the photons below 0.4 keV as diffuse emission is too much affected by galactic absorption. 
In the calculation, we assumed a galactic
hydrogen column density of $N_{H} = 5~10^{20}$cm$^2$. 
(Note that assuming a 
metallicity of 0.1 instead of 0.3 would decrease the count rate of a 2 keV
group or of 1 keV filament by   10\% or 40\% respectively in the [0.4-4] keV band).

For each set of AMR simulations we thus obtained a ``photon cube'' which was in turn
projected along the line of sight, i.e. the $x$ axis. The resulting
photon image has a pixel size equal to the apparent size of the AMR
cell seen at the simulation redshift (Table 2, step 1a); Poisson
photon noise was then added to each pixel (step 1b). This is typically
what an XMM wide angle survey would look like once the pointings have
been placed side by side, all instrumental effects corrected,
extragalactic point sources removed and the resulting image rebinned
to increase the signal (note however, this is an ideal case without
diffuse background). In a second step, we have simulated in more
detail some individual pointings, focussing on a given area of the
projected photon cube as would be seen by the circular XMM field of
view of $\sim$ 30 arcmin. 
Practically, this consists of extracting a given sky region from step 1a, rebinning it with a pixel size of 4 arcsec and then applying the vignetting function allowing for Poisson statistics in each pixel (step 2a). Finally, we have introduced the extragalactic LogN-LogS source distribution (\cite{has98}) assuming that all sources have a power law spectrum with a photon index of 2.
Images of these pointlike sources have been convolved by the mean XMM/EPIC point
spread function as a function of off-axis angle 
and randomly added to the filament sub-image together with a constant diffuse background rate of $10^{-5}$ count/s/pixel(4") (step 2b). The instrumental characteristics used in the simulations are from the ground-based calibration phase (e.g. \cite{dah99}).

\begin{table} 
\caption[]{Simulated image parameters }  

\begin{tabular}{||c| c|c|c||}
\hline \hline
step 	   &  redshift &  pixel size & image size  \\  
         	&    &  kpc \& arcsec & pixels \& arcmin  \\ \hline
1          & 0.1 &  25/h ~~~ 19" &  $952\times 1392$ ~~~ $295' \times 432'$\\
1           & 0.5 & 100/h ~~~   19"& $238 \times 348$ ~~~ $74' \times 108'$  \\ \hline
2           & 0.1 &   $\sim$ 5/h ~~~ 4"& $512 \times 512$ ~~~ $\o = 34'$ \\
2            & 0.5 &  $\sim$ 20/h ~~~ 4"&  $512 \times 512$ ~~~ $\o = 34'$ \\ \hline \hline
\end{tabular}  \\ 
\label{tab:imadim} 

In order to directly compare the filament emissivities at $z = 0.1$
and $z = 0.5$, we have rebinned the $z = 0.1$ image obtained in step
1a with a pixel size 4 times smaller than the original simulation cell
of 100/h kpc. In the calculations, we have neglected the $\sim 10\%$
difference in angular distance between the open and closed models at
$z = 0.5$.

\end{table}

\section{Results}

In order to study the effects of cosmology, distance and input physics
on the detectability by XMM of groups and filaments, we display the
simulated images of the entire filament, prior to including the
back/foreground population of X-ray point sources. We first discuss
the visibility of the $z=0.5$ case, then the $z=0.1$ one, as the
larger apparent size of the diffuse structures is likely to make their
detection more problematic, despite their proximity.

\subsection{Global images at $z = 0.5$ }

Fig. 8 shows the predicted XMM images for our three
cosmological models, namely: standard CDM, tilted CDM and
 $\Lambda$CDM. As expected, the difference between open and closed
models is striking. Conversely, the tCDM and SCDM models appear to be
very similar in terms of X-ray emissivity, at least for the group
population.  This seems quite reasonable, since both were normalized
at the 8 Mpc/h scale.  In Fig. 9, we display four
realizations of the $\Lambda$CDM model calculated under various
heating or cooling assumptions as given in Table 1, to be compared to
the $\Lambda$CDM model of Fig. 8.

\begin{figure}
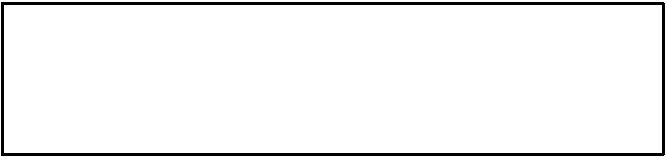

\picplace{2cm}
\caption[] {Density cube for the $z = 0.5~ \Lambda$CDM  model shown for an isosurface of $4~ 10^{-4} $ atom cm$^{-3}$  

}
\label{lCDM_75}
\end{figure}

\begin{figure}
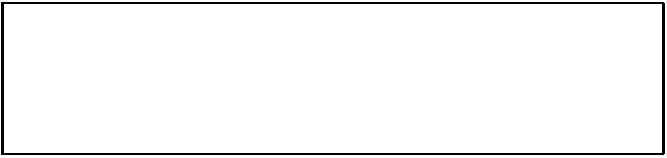

\picplace{2cm}
\caption[] {Density cube for the $ z = 0.5 ~\Lambda$CDMc model shown for an isosurface of $4~ 10^{-4} $ atom cm$^{-3}$  }

\label{lCDMc_75}
\end{figure}

Considering first the effect of cooling alone ($\Lambda$CDM ~ \& ~
$\Lambda$CDMc models), groups appear substantially smaller in spatial
extent when cooling is at work.  This is because the cooling has
reduced their pressure support, which keeps them extended.  Moreover,
the $\Lambda$CDM ~ \& ~ $\Lambda$CDMc density maps shown in
Figs. \ref{lCDM_75} \& \ref{lCDMc_75} indicate that the model with
cooling has more small clumps, but most of them do not show up in the
X-ray images: these are galaxy-sized objects, unresolved by the
simulation. In these clumps, the gas can cool down to $10^{4}$K, which
prevents most of them from emmiting in the X-ray band. In the larger
groups however, the gas in the center is still hot enough to radiate
and much denser than in the no-cooling case which makes their X-ray
profile significantly more peaked.

As discussed in Sec. 2.1, preheating raises the gas entropy prior to
gravitational collapse, preventing high central densities by the
formation of a larger core.  This is best seen when considering the
``extreme'' feed back model ($\Lambda$CDMf2 Fig. 9): clusters
are much more extended, and small objects are virtually absent. Group
central densities are significantly lower than in the $\Lambda$CDM
models, with only three clumps reaching the limiting density of $4~
10^{-4} ~$ atom cm$^{-3}$ (to compare with Figs. \ref{lCDM_75} \&
\ref{lCDMc_75}).  Some of the gas which did not succeed in collapsing
is now trapped in the filamentary structures which explains the higher
emissivity of the diffuse region compared to any other models.
Paradoxically, the average temperature in the filament is somewhat
lower in $\Lambda$CDMf2 than in $\Lambda$CDM; this is due to the absence of
shock-heating in the pre-heated filament gas.
These effects are less pronounced if mild feedback is assumed.

Finally, we considered a mixed case, combining the effects of feedback
plus cooling ($\Lambda$CDMcf1), likely to be the most realistic. This
shows very little difference from that with feedback alone (contrary
to the striking discrepancy between $\Lambda$CDM and $\Lambda$CDMc)
suggesting that here the gas stays too hot and diffuse to cool.  While
none of these models captures the detailed dynamics of real stellar
and galactic feedback, they do give an indication of the direction and
amplitude of their effects.

\subsection{Individual pointings at $z = 0.5$}

\begin{figure}
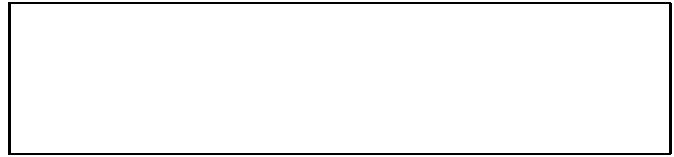

\picplace{2cm}
 \caption[] {Density cube for the $z=0.5~\Lambda$CDMcf1  model, shown for an isosurface of  $5~ 10^{-5} $ atom cm$^{-3}$   }
 
\label{lCDMcf1_10}
\end{figure}

\begin{figure}
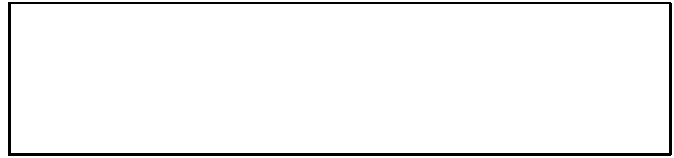

\picplace{2cm}
 \caption[] {Temperature cube for the $z=0.5~\Lambda$CDMcf1  model, shown for an isosurface of 1 keV  }
\label{temp_lCDMcf1}
\end{figure}

Figures 10 \& 11 show the two individual
pointings indicated on Fig. 9.  Although groups as small as
$10^{14}$ M$_\odot$ (or smaller depending on the model) will be visible
for integration times as short as 10 ks, the diffuse filamentary gas
seems not to be detectable even with 200 ks exposures. This is
essentially due to the very low density of the filament
(Fig. \ref{lCDMcf1_10}) and the high background level caused by source
confusion. It is interesting to note that in all models the
intra-filament gas is expected to have a temperature around 1 keV as
can be appreciated on Fig. \ref{temp_lCDMcf1}
    
\subsection{Global images at $z = 0.1$ }

The apparent size of the same filament at $z = 0.1$ is very large
($\sim 5^{o} \times 7^{o}$) and it would require more than 300
pointings to be entirely covered by XMM (assuming a spacing of 20' per
pointing to minimize the loss of sensitivity due to
vignetting). Nevertheless, as for the $z = 0.5$ models, we have
calculated the number of photons to be collected by XMM from the
entire structure for individual integration times of 200 ks. The
results are summarized in Fig. 12 for the SCDM and
$\Lambda$CDM models with or without cooling/heating. Compared with the
corresponding $z = 0.5$ images, one sees clearly the evolution of
structure in terms of merging: small groups are flowing along the
filaments to reach bigger entities located at the nodes.

\subsection{Individual pointings at $z = 0.1$}

We have performed detailed simulations of two XMM fields specifically
centered on filamentary segments (cf Fig. 12). The images are
presented in Fig. 13 and suggest that, at this redshift, very
faint details of the network will be detectable.  These may include
collapsing regions as well as pure diffuse filamentary emission.

\section{Prospects for XMM observations}

\subsection{The ``nb(photons) - Mass(groups)'' relation}

\begin{figure}
\centerline{\psfig{file=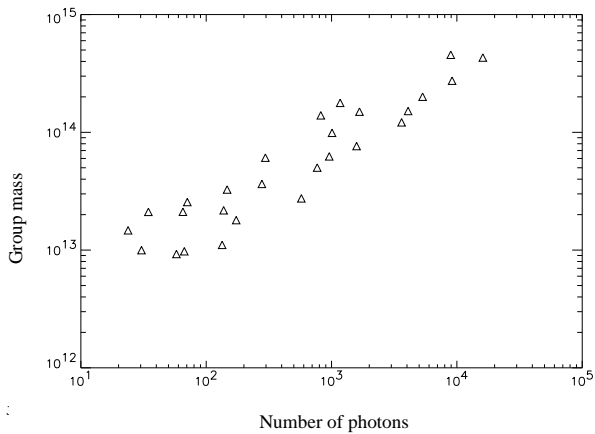,width=8cm}}
\caption[] {The relation between group mass at $z=0.5$ and the number
of photons detected in the 0.4-4 keV range from a simulated 100 ks XMM
image ($\Lambda$CDM model).
 
}
\label{mass_cluster2d}
\end{figure}

In order to get a better idea of the observability of the groups seen
in these simulated XMM images, we show in Fig.~\ref{mass_cluster2d}
the relation between the mass of a group and the number of photons
observed for the $\Lambda$CDM simulation at $z=0.5$.  The number of photons measured comes from extracting each region above a given flux
threshold from the photon cube.  The mass is calculated by summing
both the gas and the dark matter within a sphere whose mean density is
about 300 times the mean density of the universe (see \cite{bry98} for
a more complete definition).  A power law fit produces a slope of
0.59.  Simple adiabatic scaling laws would indicate a relation like $M
\sim L_X^{0.75}$, while our feedback simulations produce something
like $M \sim L_X^{0.5-0.6}$.  Note, however, that we do not directly
measure the emissivity, but count individual photons to be detected by
XMM in the [0.4-4] keV band (assuming   $ N_{H} = 5~10^{20}~cm^{-2}$).  The scatter in the relation is about
20\%.

\subsection{Shallow pointings}

With exposure times as short as 10 ks, XMM will reach a sensitivity of
$\sim 5~ 10^{-15} erg/s/cm^{2}$ in the [0.5-2] keV band for pointlike
sources. Our simulations (Fig.11) show that at this
limiting flux, objects having masses down to $10^{14}$ M$_\odot$ will
be readily detectable out to $z \sim 0.5$.  This includes all clusters and much of the group population.  As indicated in
Fig. \ref{mass_cluster2d}, groups having masses of the order of
$10^{14}$ M$_\odot$ will produce some 100 photons in 10 ks
thus, enabling simple source extent characterization as can be appreciated in Fig.~11. More generally, with 10 ks exposures, the 3$\sigma$ detection threshold for extended sources having a characteristic size of about 1 arcmin ( i.e. a $2\times$ 250 kpc core radius seen at $z \sim 0.5-1$) is around 65 photons on-axis, which is 2 times higher than for pointlike sources (\cite{val00}).

In the previous section, we demonstrated that both preheating and
cooling affects the detectability of small groups by making them
either more centrally concentrated (radiative cooling) or more
extended (preheating).  Of course it should be kept in mind that
preheating is simply our attempt to model stellar feedback.  More
sophisticated methods are being developed which attempt to account for
the multi-phase nature of the flow (\cite{wu98}; \cite{tey98}) as well
as more complete models of galaxy formation and feedback (see for
example \cite{yep97}; \cite{hul99}; \cite{mart99}).  In developing and
testing these methods, observational comparisons will play an
important role.  Therefore, a large shallow XMM survey would provide a
unique data base to study, in a statistically complete way, the
physics of groups.

\subsection{Deep pointings}

Assuming 200 ks exposures, XMM can reach a sensitivity of $\sim
10^{-16} erg/s/cm^{2}$ but will then face the generic problem of
source confusion. The limiting flux where this is expected to occur
corresponds to exposure times between 30-50 ks.  The exact level will
remain unknown until the first XMM or Chandra in-flight observations.
This is due to the lack of previous deep high resolution observations
above 2 keV (ASCA had a PSF of $\sim$ 3 arcmin). It is thus obvious
that the main hurdle for studying diffuse emission (clusters or
filaments) will be contamination by the point source population,
rather than the limited number of photons. Note that, to the contrary,
source confusion should barely be a problem for Chandra (PSF $\sim 1$ arcsec) but its much lower effective area makes the task of detecting filaments much more difficult.

An additional difficulty comes from the fact that the simulations have
a finite thickness, which is not the case of the ``observable''
universe; thus, true confusion may be even worse than presented
here because of the presence of fore/background filaments.

\subsubsection{Groups}

Assuming a 50\% efficiency for the X-ray detection algorithm in
measuring extended source fluxes and that a net minimum photon number
of 1000 is necessary for spectral studies at low energies, Fig. \ref{mass_cluster2d}
indicates that it will be possible to obtain temperature and abundance
estimates for objects having masses down to $\sim  
 10^{14} M_{\odot}$  at $z = 0.5$ (i.e.  some 2000
photons collected during an exposure time of 200 ks).  It should also
be kept in mind that we are assuming that the X-ray emission from AGN
and other galactic sources can be identified and removed from the
group X-ray emission.

Such observations will be particularly important as it will provide
new insights into the physics of low mass clusters and groups.  For
example, since supernovae are responsible for both adding energy and
metals to the intergroup medium, a measurement of the metallicity
helps to constrain the feedback history (which is even more important
in groups than in clusters).  Determinations of the baryon fraction can
also cast light on the strength of feedback and the nature of group
evolution. 

\subsubsection{Filaments}

The detection of filamentary emission (and not simply of
``supercluster gas'') is a real challenge for the next generation of X-ray satellites. With its unrivaled sensitivity and large field of view, XMM is a priori the best suited instrument. Our detailed simulations indicate that this will be very difficult at $z \sim 0.5$ whatever the assumed cosmology or heating/cooling processes, but possible at $z \sim 0.1$ for an open model associated with mild feedback ($\Lambda$CDMf1, Sec. 4.4).

More generally, in this work, we have tried to gauge the effect of
uncertainties in our understanding of the physics of filament
(i.e. cooling and feedback).  We have shown that cooling alone tends
to decrease somewhat
the luminosity of filaments.  This occurs because gas cools
into many small clumps, reaching temperatures which are too low to
radiate in the X-ray band.  However, this also produces groups and
clusters which are much brighter than observations indicate (as
reflected in the mismatched $L_X-T$ relation).  

On the other hand, in the simulation with the extreme form of
preheating ($\Lambda$CDMf2), small groups are prevented from forming at all
(in a more realistic simulation, these groups would form but then
expel nearly all of their gas due to supernovae in galaxies).  This
causes more gas to end up in the filaments, improving their
visibility.  

The case of mild feedback, which satisfies all the observational
constraints is intermediate in terms of filament emission.  Feedback and
cooling ($\Lambda$CDMcf1, the most likely model) appears very
similar to the model with feedback alone ($\Lambda$CDMf1) in terms of emissivity. Thus, our prediction that filamentary emission will be
detectable at $z \sim 0.1$ is likely to be robust. 
Note also that the variation due to changes in input physics is much
smaller for filaments than for groups and
filaments, which are more sensitive to the nature and strength of the
feedback.  

The number of photons received from  the filament detectable in Fig.
13 (bottom) is $\sim$ 21000 for an integration time of 200 ks.
This is, in principle, 10 times more than strictly necessary for
measuring the temperature of the feature. However, the filament
photons are spread over the entire detector and severely contaminated
by the point source population (resolved or not). With the present
analysis techniques, it is not possible to extract a workable spectrum.

\subsection{Technical remark on filtering techniques}

Multi-resolution filtering techniques give very good results for
enhancing the signal from point source and cluster type emission as
well as assessing the significance of the detected objects (cf Fig.
10 \& 11).
However, our preliminary processing shows that gaussian filtering is
better adapted for enhancing very diffuse structures having a size
comparable to that of the detector. In a forthcoming paper (Valtchanov
et al 2000), we shall investigate further source detection and extraction techniques and the possibility of obtaining spectroscopic information from diffuse filamentary regions.

\section{Conclusion. How can the cosmology be constrained?}

We have presented detailed  predictions as to the visibility of the
cosmic structure by XMM using high resolution hydrodynamical
simulations.  Special attention has been paid to  physical processes
--- in addition to pure gravitational infall and shocks --- which can
affect the X-ray emissivity.  The discussion was focussed on the
lowest density objects, i.e., small groups of galaxies and cosmic
filaments.

One of the main conclusion is that the entire (or much of) the low
mass group population will be visible with modest exposure times out
to a redshift of one half, depending on the adopted cosmology and
details of the physics. More specifically, for open universes, we have
investigated two effects (cooling and feedback) using simplified
prescriptions.  We now place them in a more general context: (i)
Feedback from galaxies, such as supernova driven galactic winds, is
expected to be particularly efficient prior to group formation since
the rather small masses of groups makes them more sensitive to the
entropy of the gas from which they form.  Energy and matter input to
the ICM occurs throughout the lifetime of the group, but probably but
with less impact on their luminosities than with the preheating
adopted here (\cite{met94}); (ii) Further, because of their lower
temperatures, groups are to be also strongly affected by cooling. This
process is essentially multi-phase and because of the
condensation of cooling clouds, the remaining hot gas has an X-ray
luminosity strongly reduced compared to what would be expected from an
equivalent adiabatic group; the X-ray mass fraction is also
significantly influenced by this multi-phase evolution.  As we have
emphasized in this work, neither of these effects are currently well
understood or modelled.

The XMM group sample should then provide precious insights into these
various mechanisms. However, a comprehensive analysis of the group
population should necessarily also involve a combined optical study,
since galaxy distributions and velocities are also key parameters in
understanding the history of groups (an example is the so-called
$\beta$ problem, e.g. \cite{gerb94}). 

While less numerous and significantly less affected by cooling or
feedback, large clusters --- the deepest potential wells in the
universe --- play an important role in our current cosmological
constraints.  For example, their number density provides the best
estimate of the amplitude of the primordial density fluctuations at
scales around 8$h^{-1}$ Mpc (e.g. \cite{via96}).  Similarly, their
evolution in redshift has been suggested by many authors
as a way to measure the matter density of the
universe.  However, a convincing measurement of this has
proven difficult due to the uncertain relation between a cluster's
mass and its X-ray luminosity.  In fact, this has driven many workers
to use the temperature, which is more difficult to measure, in
addition to the luminosity since the former is more closely related to
the mass (e.g. \cite{ouk97}).
Since much of the uncertainty in understanding cluster luminosities
stems from feedback, the  proposed observations for the low mass sample (along with improved modeling) will be able to cast light on such things as the $L_X-T$ relation, and so provide much tighter cosmological constraints.
 
Even more interesting, however, is the idea of using the spatial
distribution of groups and filaments themselves to constrain
cosmology.  Even a cursory glance at Fig. 8 shows that the
width, length and distribution of filaments is sensitive to cosmology
(in fact, the dependence at $z=0$ is almost entirely on the amplitude
of the power spectrum).  The difficulty is extracting this information
from the more realistic observations shown in later figures.  One
possibility is to use the location of small groups to trace out the
filaments and apply topological measurements, such as the genus
statistic.  This has been done in the case of galaxies.  A difficulty
with this and other approaches is simply the weak signal of the
filaments.  In this paper we have shown that they are detectable with
XMM, but confusion will be an issue.  A promising approach to increase
the signal-to-noise ratio is to combine multiple measurement
techniques by cross-correlating galaxy surveys, X-ray maps and perhaps
even microwave maps, to use the expected Sunyaev-Zel'dovich signal.
This will help to eliminate the confusion caused by unrelated (and
hence uncorrelated) sources. In the optical band, weak lensing
analysis techniques are rapidly improving and it may soon become
possible to measure the direct gravitational signal from the mass
associated with
filaments (preferentially located between two clusters). Having
gained a better understanding of feedback and cooling mechanisms from
the study of groups, we may then be in a position to interpret the
observed X-ray emissivity in terms of density, temperature and
abundances and hence infer the baryon fraction in filaments.\\

{\em Acknowledgements}:  \\
Support for GLB was provided by NASA through Hubble Fellowship grant
HF-01104.01-98A awarded by the Space Telescope Science Institute,
which is operated by the Associate of Universities for Research in
Astronomy, Inc., for NASA under contract NAS 5-26555.  We thank
the National Center for Supercomputing Applications for providing
computational resources for some of the simulations carried in this
work.  We acknowledge useful conversations with Alexandre Refregier.
We are grateful to H. Brunner and G. Hasinger for detailed information used in the XMM simulations. \\

{}

\vspace{5cm}

\noindent

\begin{center}
 {\bf Colour Fig. 8 to 13  and corresponding captions \\ \mbox{are available at:} http://www.mit.edu/$\sim$gbryan/xmm\_preview.html}

\end{center}

\begin{thebibliography}{}

\bibitem[Allen \& Fabian 1998]{all98}
Allen S.W., Fabian A.C., 1998, MNRAS 297, L63-L68

\bibitem[Arnaud \& Evrard 1999]{AE99}
Arnaud M., Evrard A., 1999, MNRAS 305, 631

\bibitem[Balogh, Babul \& Patton 1999]{bal99}
Balogh M.L., Babul A. \& Patton D.R. 1999, MNRAS, in press

\bibitem[Bond et al 1996]{bon96}
Bond R., Kofman L., Pogosyan D., 1996, Nature, 380, 603 


\bibitem[Briel \& Henry 1995]{bri95}
Briel, U.G. \& Henry, J.P. 1995, A\&A, 302, L9

\bibitem[Bryan et al. 1995]{bry95}
Bryan G.L., Norman M.L., Stone J.M., Cen R., Ostriker J.P., 1995,
Comp. Phys. Comm. 89, 149-168

\bibitem[Bryan \& Norman 1997]{bry97}
Bryan G.L., Norman M.L., 1997, in ``Computational Astrophysics'',
Proc. 12th Kingston Conference, Halifax, Oct. 1996, eds. D. Clarke \&
M. West (PASP), p. 363-368

\bibitem[Bryan \& Norman 1998]{bry98}
Bryan G.L. \& Norman M.L., 1998, ApJ, 495, 80

\bibitem[Cen \& Ostriker 1999]{cen99}
Cen R. \& Ostriker, J.P, 1999, ApJ, 514, 1

\bibitem[Colella \& Woodward 1984]{col84}
Colella P., Woodward P.R., 1984, J. Comp. Phys. 54, 174


\bibitem[Dahlem \& Shartel 1999]{dah99}
Dahlem M. \& Schartel N. 1999, {\em The XMM Users} Hanbook,http://astro.estec.esa.nl/XMM/user/uhb\_top.html 


\bibitem[Gerbal et al 1994]{gerb94}
Gerbal D., Durret F., Lachièze-Rey M., 1994, A\& A 288, 746

\bibitem[Hasinger et al 1998]{has98}
Hasinger G., Burg R., Giacconi R., Schmidt M., Tr\"umper J., Zamorani G.,  1998,  A\& A 329, 482

\bibitem[Hultman \& Pharasyn 1999]{hul99}
Hultman, J. \& Pharasyn A. 1999, A\&A, 347, 769

\bibitem[Kaiser 1991]{kai91}
Kaiser N. 1991, ApJ, 383, 104

\bibitem[Kull \& B\"ohringer 1999]{kul99}
Kull, A. \& B\"orhigner, H. 1999, A\&A, 341, 23

\bibitem[Markevitch 1999]{mar99}
Markevitch, M. 1999, ApJ, 522, L13

\bibitem[Martin 1999]{mart99}
Martin, C.L. 1999, ApJ, 513, 156

\bibitem[Metzler \& Evrard 1994]{met94}
Metzler, C.A., Evrard, A.E. 1994, ApJ, 437, 564

\bibitem[Miralda-Escud\'e et al. 1996]{mir96}
Miralda-Escud\'e, J., Cen, R., Ostriker, J.P. \& Rauch, M. 1996,
ApJ, 471, 582

\bibitem[Mulchaey \& Zabludoff 1998]{mul98}
Mulchaey, J.S. \& Zabludoff, A.I. 1998, ApJ, 496, 73

\bibitem[Mushotzky \& Scharf 1997]{MS97}
Mushotzky R.F., Scharf C.A., 1997, ApJ Letters 482, L13
 
\bibitem[Navarro, Frenk \& White 1996]{nav96}
Navarro J.F., Frenk. C.S., White S.D.M. 1996, ApJ, 462, 563

\bibitem[Norman \& Bryan 1999]{nor99}
Norman M.L., Bryan G.L., 1999, to appear in ``Numerical Astrophysics
1998'', eds. S. Miyama \& K. Tomisaka, Tokyo, March 10-13, 1998

\bibitem[Oukbir \& Blanchard 1997]{ouk97}
Oukbir, J. \& Blanchard, A. 1997, A\&A, 317, 13
 
\bibitem[Peebles 1980]{pee80}
Peebles, P.J.E., 1980 {\em Large Scale Structuree of the Universe} (Princeton University Press) 
\bibitem[Reichart, Castander \& Nichol 1999]{rei99a}
Reichart, D.E., Castander, F.J., Nichol, R.C. 1999, ApJ, 516

\bibitem[Reichart et al. 1999]{rei99b}
Reichart, D.E., Nichol, R.C., Castander, F.J., Burke, D.J., Romer,
A.K., Holden, B.P., Collins, C.A., Ulmer, M.P. 1999, ApJ, 518, 521

\bibitem[Renzini 1997]{ren97}
Renzini, A. 1997, ApJ, 488, 35

\bibitem[Rosati et al. 1998]{ros98}
Rosati, P. Della Ceca, R., Norman, C., Giacconi, R. 1998, ApJ, 492,
21L

\bibitem[Sarazin \& White 1987]{sar87}
Sarazin C.L., White R.E. 1987, ApJ, 320, 32

\bibitem[Starck \& Pierre 1998]{sta98}
Starck J.-L., Pierre M., 1998, A\& A Sup., 128, 397

\bibitem[Teyssier, Chieze \& Alimi 1998]{tey98} 
Teyssier, R., Chieze, J.P., Alimi, J.M. 1998, in ``Untangling Coma
Berenices: A New Vision of an Old Cluster'', Proceedings of the
meeting held in Marseilles (France), June 17-20, 1997, Eds.: Mazure,
A., Casoli F., Durret F. , Gerbal D., Word Scientific Publishing Co
Pte Ltd, p 149.


\bibitem[Valtchanov et al 2000]{val00}
Valtchanov I., Gastaud R., Pierre M., Starck J.-L.,  2000 A\&A in preparation

\bibitem[Viana \& Liddle 1996]{via96}
Viana, P.T.P., Liddle, A. 1996, MNRAS, 281, 531L

\bibitem[Wang et al. 1997]{wan97}
Wang, Q. Daniel, Connolly, A., Brunner, R. ApJ, 487, 13L

\bibitem[Wu, Fabian \& Nulsen 1998]{wu98}
Wu, K.K.S., Fabian, A.C., Nulsen, P.E.J. 1998, MNRAS, 301, 20L

\bibitem[Yepes et al. 1997]{yep97}
Yepes, G., Kates, R., Khokhlov, A., Klypin, A. 1997, MNRAS, 284, 235

\bibitem[Zel'dovich 1970]{zel70}
Zel'dovich, Ya. B. 1970, Astrofizika 6,319,

\end{thebibliography}
\end{document}